\documentclass[twocolumn,showpacs,preprintnumbers,amsmath,amssymb]{revtex4}

\usepackage{graphicx}

\begin{document}
\title{Observation of scale invariance and universality in two-dimensional Bose gases}
\author{Chen-Lung Hung}
\author{Xibo Zhang}
\author{Nathan Gemelke\footnote{Current address: Department of Physics, The Pennsylvania State University, University Park, Pennsylvania 16802, USA}}
\author{Cheng Chin}
\affiliation{The James Franck Institute and Department of Physics, \\ The University of Chicago, Chicago, IL 60637, USA}
\date{\today}

\begin{abstract}
The collective behavior of a many-body system near a continuous phase transition is insensitive to the details of its microscopic physics\cite{Stanley99}. Characteristic features near the phase transition are that the thermodynamic observables follow generalized scaling laws\cite{Stanley99}. The Berezinskii-Kosterlitz-Thouless (BKT) phase transition\cite{Berezinskii72,Kosterlitz73} in two-dimensional (2D) Bose gases presents a particularly interesting case because the marginal dimensionality and intrinsic scaling symmetry\cite{Pitaevskii97} result in a broad fluctuation regime which manifests itself in an extended range of universal scaling behavior. Studies on BKT transition in cold atoms have stimulated great interest in recent years\cite{Hadzibabic06,Kruger07,Schweikhard07,Clade09,Hadzibabic08, Tung10}, clear demonstration of a critical behavior near the phase transition, however, has remained an elusive goal. Here we report the observation of a scale-invariant, universal behavior of 2D gases through in-situ density and density fluctuation measurements at different temperatures and interaction strengths. The extracted thermodynamic functions confirm a wide universal region near the BKT phase transition, provide a sensitive test to the universality prediction by classical-field theory \cite{Prokof'ev01,Prokof'ev02} and quantum Monte Carlo (MC) calculations\cite{Holzmann10}, and point toward growing density-density correlations in the fluctuation region. Our assay raises new perspectives to explore further universal phenomena in the realm of classical and quantum critical physics. 
\end{abstract}


\maketitle

\begin{figure}[t]
\includegraphics[width=88 mm]{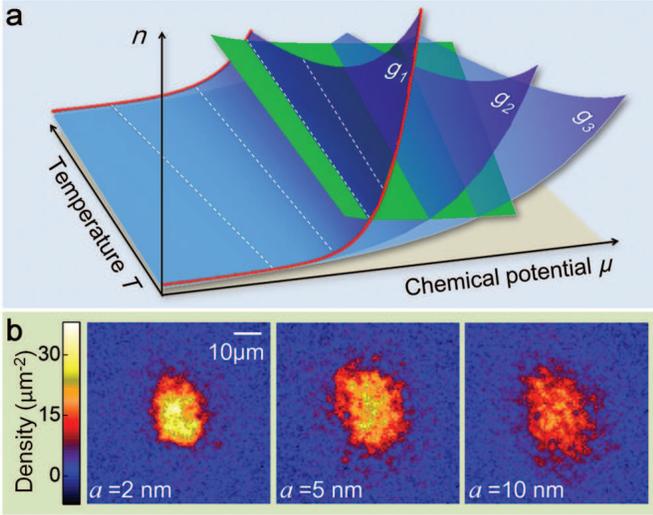}
\caption{Illustration of scale invariance and universality in 2D quantum gases. \textbf{a,} Scale invariance links any thermodynamic observable at different $\mu$ and $T$ via a simple power-law scaling. In a 2D Bose gas with coupling constant $g \ll 1$, atomic density $n$ measured at different temperatures (red lines) can be scaled through constant $\mu/T$ and $n/T$ contours (dashed lines). Near the BKT phase transition boundary (green plane), systems with different $g=g_1, g_2...$ (blue planes) scale universally. 
\textbf{b,} \textit{In situ} density measurements of trapped 2D gases provide crucial information to test the hypotheses of scale invariance and universality. Sample images at different scattering lengths $a$ are obtained from single shot.} \label{fig1}
\end{figure}

In 2D Bose gases, critical behavior develops in the BKT transition regime, where an ordered phase with finite-ranged coherence
competes with thermal fluctuations and induces a continuous phase
transition from normal gas to superfluid with quasi-long range
order\cite{Kosterlitz73}. In this fluctuation region, a universal and scale-invariant description for the system is expected through the power-law scaling of thermodynamic quantities with respect to the coupling strength
and a characteristic length scale\cite{Prokof'ev02,Holzmann07}, e.g., thermal de Broglie wavelength (Fig.~1a). For weakly interacting gases at finite temperatures, in particular, the scale invariance prevails over the normal, fluctuation, and superfluid regions because of the density-independent coupling constant\cite{Petrov00} and the symmetry of underlying Hamiltonian\cite{Pitaevskii97}.

In this letter, we verify the scale invariance and universality of
interacting 2D Bose gases, and identify BKT critical points. We test
scale invariance of \textit{in situ} density and density
fluctuations of $^{133}$Cs 2D gases at various temperatures. We
study the universality near the BKT transition by tuning the atomic
scattering length using a magnetic Feshbach resonance\cite{Chin10}
and observing a universal scaling behavior of the equation of state
and the quasi-condensate density. Finally, by comparing the local
density fluctuations and the compressibility derived from the
density profiles, we provide strong evidence of a growing density-density correlation in the fluctuation regime.

We begin the experiment by loading a nearly pure $^{133}$Cs Bose
condensate of $N=2\times 10^4$ atoms into a single pancake-like
optical potential with strong confinement in the vertical ($z$-)
direction and weak confinement in the horizontal ($r$-)
direction\cite{Gemelke09,Hung10}. The trapping potential, $V(r,z)= m\omega_r^2 r^2/2 + m\omega_z^2 z^2/2$, has mean harmonic trapping frequencies $\omega_r = 2 \pi \times 10$~Hz and $\omega_z = 2\pi \times 1900$~Hz. Here, $r$ denotes the radial distance to the trap center and $m$ is the cesium atomic mass. In this trap, the gas reaches temperatures as low as $T = 15$~nK and moderate peak chemical potential $\mu_0 < k_B T$. The ratio $\hbar \omega_z/\mu_0 > \hbar \omega_z/k_B T \sim 6$ indicates that the sample is deeply in the 2D regime with $<1 \%$ population in the vertical excited states. Here, $\hbar=h/2\pi$, $h$ is the Planck constant, and $k_B$ is the Boltzmann constant. The 2D
coupling constant is evaluated according to $g = \sqrt{8 \pi}a/l_z$
\cite{Petrov00}, where $a$ is the atomic scattering length
and $l_z=200$~nm is the vertical harmonic oscillator length. We
control the scattering length $a$ in the range of $2$~$\sim
10$~nm, resulting in weak coupling strengths $g=0.05
\sim 0.26$. Here, the density-dependent correction to $g$\cite{Petrov00,Mora09} is expected to be small and negligible ($<2\%$).

We obtain \textit{in situ} density distributions of 2D gases by
performing absorption imaging perpendicular to the horizontal plane
with a commercial microscope objective and a CCD camera\cite{Hung10} (see Fig.~1b for sample images). About 50 images are collected for each experiment condition, and the average density $n$ and the density variance $\delta n^2$ are evaluated pixel-wise (see Methods). We obtain the radial density $n(r)$ and variance $\delta n^2 (r)$ profiles (Fig.~2 insets) by accounting for the cloud anisotropy and performing azimuthal averaging\cite{Gemelke09}.

We obtain the equation of state $n(\mu,T)$ from the averaged density
profile by assigning a local chemical potential $\mu(r) = \mu_0 -
V(r,0)$ to each point according to the local density approximation. Both $T$ and $\mu_0$ can be determined from the low density wing where the sample is assumed normal and the density profile can be fit to a mean-field formula $n(\mu,T) = -\lambda_{dB}^{-2}  \ln[1- \exp(\mu/k_B T- g n \lambda_{dB}^2/\pi)]$\cite{Hadzibabic08}, where $\lambda_{dB}=h/\sqrt{2\pi m k_B T}$ is the thermal de Broglie wavelength.

\begin{figure}[t]
\includegraphics[width=88 mm]{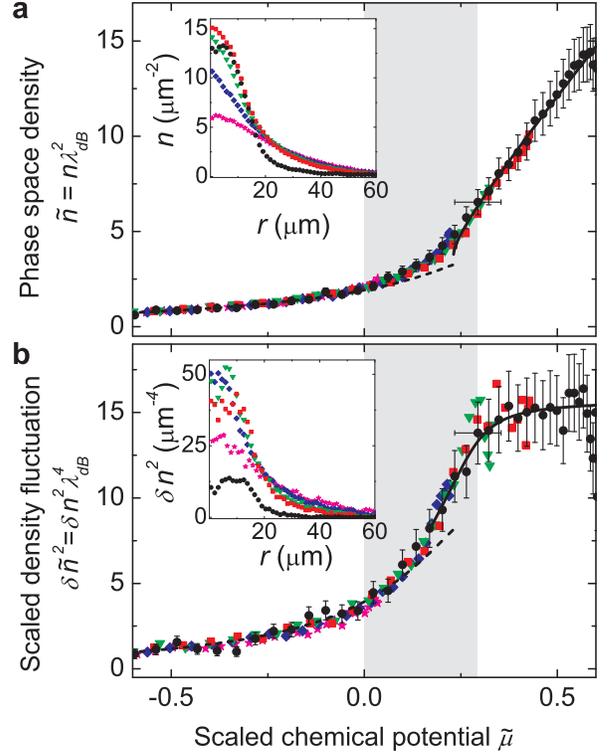}
\caption{Scale invariance of density and its fluctuation. \textbf{a}, Scaled density (phase space density) $\tilde{n} =n\lambda_{dB}^2$ as a function of the scaled chemical potential $\tilde{\mu} =\mu/k_BT$ measured at five different temperatures: $T=21$~nK (black circles), 37~nK (red squares), 42~nK (green triangles), 49~nK (blue diamonds), and 60~nK (magenta stars), and coupling strength $g=0.26$. Mean-field expectations for normal gas (dashed line) and superfluid (solid line) are shown for comparison. Inset shows the radial density profiles before scaling. \textbf{b}, Scaled fluctuation $\delta \tilde{n}^2=\delta n^2\lambda_{dB}^4$ at different temperatures. Dashed line is the mean-field calculation based on the fluctuation-dissipation theorem\cite{Huang63}. Solid line is an empirical fit to the crossover feature from which the critical chemical potential $\tilde{\mu}_c$ is determined. Inset shows the radial fluctuation profiles before scaling. The shaded area marks the fluctuation region $0<\tilde{\mu}<\tilde{\mu}_c$. Error bars show standard deviation of the measurement.} \label{fig2}
\end{figure}

We confirm the scale invariance of a 2D gas by first introducing the dimensionless, scaled forms of density $\tilde{n}=n\lambda_{dB}^2$ (phase space density), fluctuation $\delta \tilde{n}^2 = \delta n^2 \lambda^4_{dB}$, and chemical potential $\tilde{\mu}=\mu/k_B T$, and showing that the equation of state and the fluctuation satisfy the following forms:
\begin{eqnarray}
\tilde{n} &=& F(\tilde{\mu})\\
\delta \tilde{n}^2 &=& G(\tilde{\mu}),
\end{eqnarray}
where $F$ and $G$ are generic functions. This suggests both energy and length scales are set solely by the thermal energy and the de Broglie wavelength, respectively. An example at $g=0.26$ ($a=10$~nm) is shown in Fig.~2. Here we show that while the original density and fluctuation profiles are temperature dependent (see Fig.~2 insets), all profiles collapse to a single curve in the scaled units. At negative chemical potential $\tilde{\mu}<0$, the system is normal and can be described by a mean-field model (dashed lines). In the range of $0<\tilde{\mu}<0.3$, the system enters the fluctuation regime and deviation from the mean-field calculation becomes evident. Crossing from normal gas to this regime, however, we do not observe a sharp transition feature in the equation of state. At even higher $\tilde{\mu} > 0.3$, the system becomes a superfluid and the density closely follows a mean-field prediction\cite{Prokof'ev02} $\tilde{n} = 2\pi \tilde{\mu}/g + \ln(2\tilde{n}g/\pi-2\tilde{\mu})$. We notice that the mean-field theory in the superfluid limit also cannot accurately describe the system in the fluctuation regime. Transition into the BKT superfluid phase is most easily seen in the scaled fluctuation $\delta\tilde{n}^2$, which crosses over to a nearly constant value due to the suppression of fluctuation in the superfluid regime\cite{Simula08}. In the density profile $\tilde{n}$, a corresponding transition feature can be found when one computes the derivative $\partial \tilde{n}/\partial \tilde{\mu}$, i.e., the scaled compressibility $\tilde{\kappa}$, as suggested by the fluctuation-dissipation theorem discussed in later paragraphs and Fig.~4. Finally, our measurement suggests that the validity of scale invariance extends to all thermal, fluctuation and superfluid regimes, a special feature for weakly-interacting 2D gases\cite{Pitaevskii97} which underlies the analysis of a recent experiment\cite{Rath10}.

We associate the crossover feature in the density fluctuations $\delta\tilde{n}^2$ and the scaled compressibility $\tilde{\kappa}$ with the BKT transition\cite{Simula08, Holzmann08}. To estimate the location of the transition point, we apply an empirical fit to this feature and determine the critical chemical potential $\tilde{\mu}_c$ and the critical phase space density $\tilde{n}_c$ (see full Methods). Results at different $g$ in the range of 0.05 to 0.26 are shown in Fig.~3c-d and compared to the theoretical prediction of $\tilde{n}_c = \ln (\xi/g)$ and $\tilde{\mu}_c = (g/\pi) \ln (\xi_{\mu}/g)$\cite{Popov83}, where $\xi=380$ and $\xi_{\mu}=13.2$ are determined from a classical-field MC calculation\cite{Prokof'ev01}. Our results show good agreement with the theory, apart from a potential systematic error from the choice of the fit function, which can account for a down shift of $10\%$ in the fit values of $\tilde{\mu}_c$ and $\tilde{n}_c$.

Further comparison between profiles at different interaction strengths allows us to test the universality of 2D Bose gases. Sufficiently close to the BKT critical point with $|\tilde{\mu}-\tilde{\mu}_c| < g$, one expects the phase space density shows a universal behavior\cite{Prokof'ev02},
\begin{equation}
\tilde{n}-\tilde{n}_c = H(\frac{\tilde{\mu}-\tilde{\mu}_c}{g}),
\end{equation}
where $H$ is a generic function. Here, density and chemical potential are offset from the critical values $\tilde{n}_c$ and $\tilde{\mu}_c$, which remove the non-universal dependence on the microscopic details of the interaction\cite{Holzmann07,Prokof'ev02}.

To test the universality hypothesis, we rescale $\tilde{\mu}$ to $\tilde{\mu}/g$ and look for critical values $\tilde{n}_c$ and $\tilde{\mu}_c$ such that the equations of state at all values of $g$ display a universal curve in the phase transition regime (see full Methods). Indeed, we find that all rescaled profiles can collapse to a single curve in the fluctuation region $-1 < (\tilde{\mu}-\tilde{\mu}_c)/g < 0$ and remain overlapped in an extended range of $|\tilde{\mu}-\tilde{\mu}_c|/g \le 2$ (see Fig.~3a), which contrasts the very different equations of state $\tilde{n}(\tilde{\mu})$ at various $g$ shown in the inset of Fig. 3a. Our result closely follows the classical-field prediction\cite{Prokof'ev02} and quantum MC calculations\cite{Holzmann10} assuming strictly 2D mean-field contribution, and the fitting parameters: critical density $\tilde{n}_{c}$ and chemical potential $\tilde{\mu}_{c}$ show proper dependence on $g$ and are in fair agreement with the theory prediction\cite {Prokof'ev01} (see Fig.~3c-d). We emphasize that critical values determined from the density fluctuations (see Fig.~3c-d) match well with those determined from the universal behavior, indicating that universality is a powerful tool to determine the critical point from a continuous and smooth density profile. Similar agreement with the theory on the critical densities has also been reported based on different experiment techniques\cite{Kruger07,Clade09,Tung10}.

\begin{figure}[t]
\includegraphics[width=88 mm]{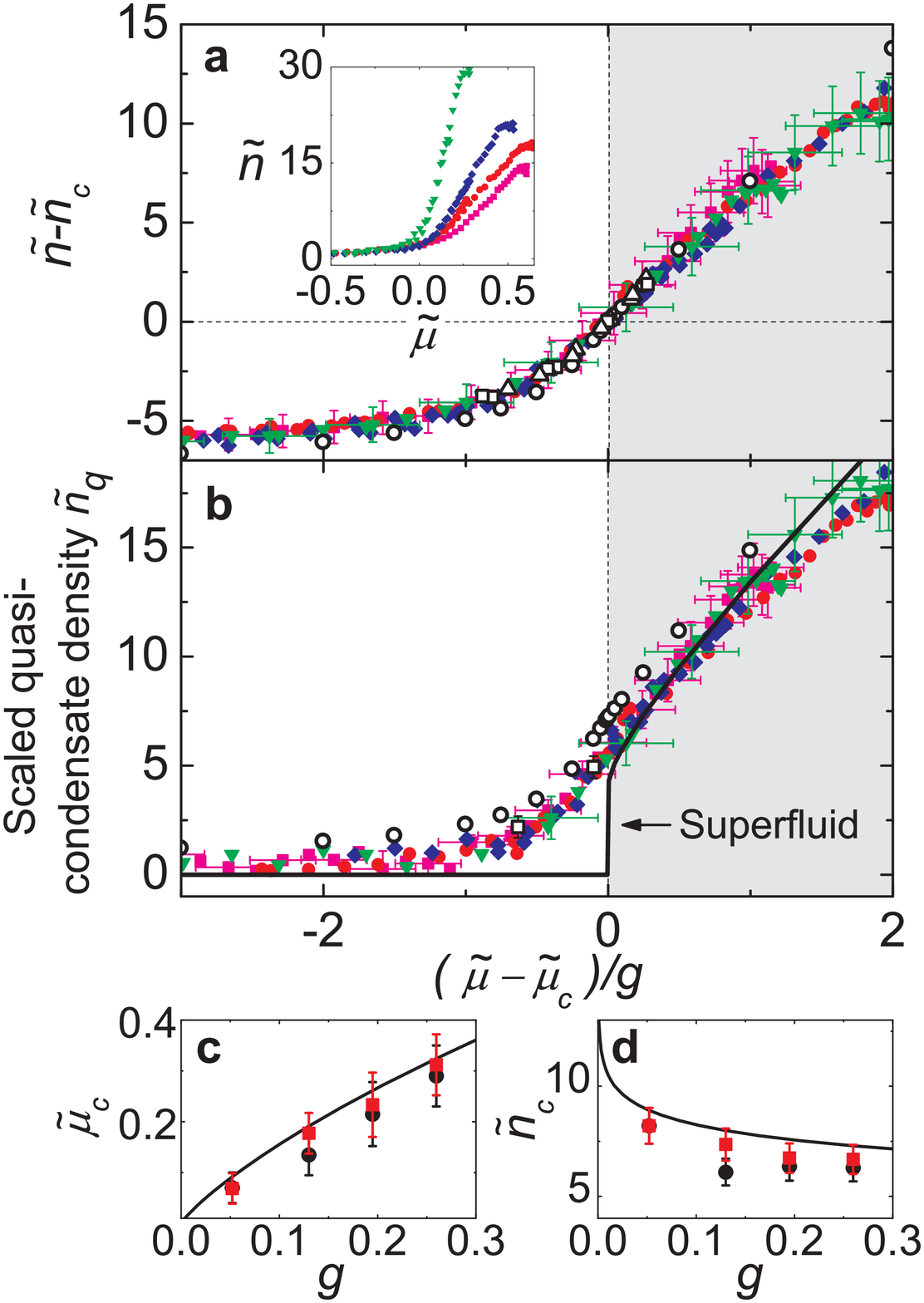}
\caption{Universal behavior near the BKT critical point. \textbf{a}, Rescaled density profiles $\tilde{n}-\tilde{n}_c$ measured at various coupling strengths, $g=0.05$ (green triangles), 0.13 (blue diamonds), 0.19 (red circles), and 0.26 (magenta squares). Inset shows the original equations of state $\tilde{n}(\tilde{\mu})$. \textbf{b}, scaled quasi-condensate density $\tilde{n}_q=\sqrt{\tilde{n}^2-\delta \tilde{n}^2}$ at different interaction strengths. In both plots, MC calculations from Ref.~\cite{Prokof'ev02} (open circles) and Ref.~\cite{Holzmann10} (\textbf{a}, open squares for $g=0.07$ and open triangles for $g=0.14$; \textbf{b}, open squares) are plotted for comparison. The shaded area marks the superfluid regime and the solid line in \textbf{b} shows the superfluid phase space density calculation\cite{Prokof'ev02}. \textbf{c} and \textbf{d}, critical values $\tilde{\mu}_c$ and $\tilde{n}_c$ determined from the following methods: universal scaling as shown in \textbf{a} (see full Methods, red squares), density fluctuation crossover (see text, black circles), and MC calculation from Ref.~\cite{Prokof'ev01} (solid line). Experiment values coincide at $g=0.05$ identically, as a result of our analysis (see full Methods). Error bars show the standard deviation of the measurement.} \label{fig3}\end{figure}

Further universal features near the phase transition can be revealed in the growth of the quasi-condensate (QC) density $n_q =\sqrt{n^2-\delta n^2}$ across the phase transition\cite{Prokof'ev01,Prokof'ev02,Bisset09}. QC is a measure of the non-thermal population in a degenerate Bose gas. A finite QC density does not necessarily imply superfluidity, but can be responsible for a non-Gaussian distribution observed in momentum space\cite{Clade09}. QC is predicted to be universal near the critical point following\cite{Prokof'ev02}

\begin{equation}
\tilde{n}_q = Q(\frac{\tilde{\mu}-\tilde{\mu}_c}{g}),
\end{equation}

\noindent where $Q$ is a generic function and $\tilde{n}_q=n_q \lambda_{dB}^2$.

We employ both of our density and fluctuation measurements to evaluate $\tilde{n}_q$ at various $g$. Adopting $\tilde{\mu}_c$ determined from the universal behavior of the density profile, we immediately find that all measurements collapse to a single curve in the range of $|\tilde{\mu}-\tilde{\mu}_c|/g \leq 2$ with apparent growth of QC density entering the fluctuation region (Fig.~3b). The generic function $Q$ we determined is in good agreement with the classical-field\cite{Prokof'ev02} and quantum MC\cite{Holzmann10} calculations with no fitting parameters. Both our density and fluctuation measurements show universal behaviors throughout the fluctuation region where a mean-field description fails and confirm universality in a 2D Bose gas near the BKT phase transition\cite{Prokof'ev02,Holzmann10}.

The generic functions we described in the previous paragraphs offer new avenues to investigate the critical behavior of the 2D gas. Following the framework of scale invariance, we compare the dimensionless compressibility $\tilde{\kappa} =
\partial \tilde{n} / \partial \tilde{\mu} = F'(\tilde{\mu})$ and the fluctuation
$\delta \tilde{n}^2 = G(\tilde{\mu})$ extracted from the measurements at $g=0.05$ and $0.26$ (see Fig. 4). In the normal gas regime at low phase space density ($G(\tilde{\mu}), F'(\tilde{\mu})<3$), a simple equality $G=F'$ is observed. This result is consistent with the fluctuation-dissipation theorem (FDT) for a classical grand canonical ensemble\cite{Huang63}, which gives $k_BT \frac{\partial N}{\partial \mu} = \delta N^2$, where $N$ is the particle number in a detection cell. In the fluctuation and the superfluid regimes at higher phase space density, our measurement shows that density fluctuations drop below the compressibility $G<F'$.

\begin{figure}[t]
\includegraphics[width=88 mm]{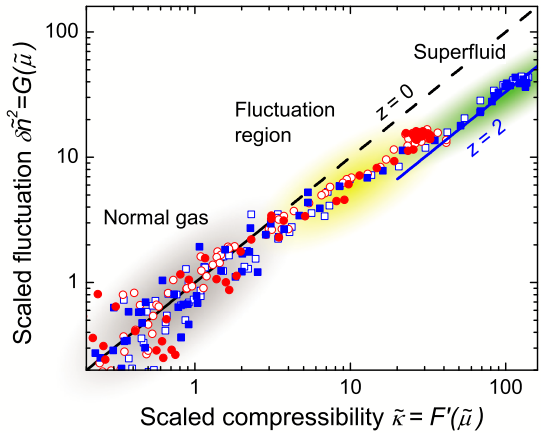}
\caption{Fluctuation versus compressibility. Scaled compressibility $\tilde{\kappa}=F'(\tilde{\mu})$ and scaled density fluctuation $\delta \tilde{n}^2=G(\tilde{\mu})$ are derived from measurements at two interaction strengths, $g=0.05$ (squares) and $g=0.26$ (circles), each containing two different temperatures between 20 and 40~nK (solid and open symbols, respectively). Diagonal line shows the expectation of $G = F'$ in the normal gas region. Solid line shows suppressed fluctuation $G = F'/ (1+z)$ with $z=2$. The color shaded areas mark the regimes of normal (left), fluctuation (middle), and superfluid (right).} \label{fig4}
\end{figure}

Natural explanations for the observed deviation include non-vanishing dynamic density susceptibility at low temperature\cite{Kubo66} and the emergence of correlations in the fluctuation region\cite{Toda92}. While the former explanation is outside the scope of this article, we show that the correlation alone can explain our observation. Including correlation, the compressibility conforms to\cite{Zhou09,Toda92}

\begin{eqnarray}
\tilde{\kappa}(\mathbf{r}) &=& \lambda_{dB}^{-2} \int \langle \delta \tilde{n}(\mathbf{r})  \delta \tilde{n} (\mathbf{r}+\mathbf{r'})\rangle d^2r' \\
&=&\delta \tilde{n}^2(\mathbf{r})(1+ z),
\end{eqnarray}
where $\langle ... \rangle$ denotes ensemble average and $z=\frac{1+ n(\mathbf{r}) \int [g^{(2)}(\mathbf{r},\mathbf{r}+\mathbf{r'})-1]d^2r'}{1+ n(\mathbf{r}) \int_v [g^{(2)}(\mathbf{r},\mathbf{r}+\mathbf{r'})-1]d^2r'}-1$ is the relative strength of correlation to local fluctuation $\delta \tilde{n}^2$ \cite{Toda92}. Here $g^{(2)}$ is the normalized second-order correlation function\cite{Naraschewski99} and $v$ denotes the effective area of the resolution limited spot. When the sample is uncorrelated, we have $z=0$; non-zero $z$ suggests finite correlations in the sample. In the fluctuation region shown in Fig.~4, observing a lower fluctuation than would be indicated by the compressibility, with $z$ approaching 2, suggests that the correlation length approaches or even exceeds our imaging cell dimension $\sqrt{v} \sim 2~\mu$m. This observation is in agreement with the expected growth of correlation when the system enters the fluctuation region. Similar length scales were also observed in the first-order coherence near the BKT phase transition using an interferometric method\cite{Clade09} and near the superfluid phase transition in three dimensions\cite{Donner07}.

In summary, based on \textit{in situ} density measurements at different chemical potential, temperature, and scattering length, we have explored and confirmed the global scale invariance of a weakly-interacting 2D gas, as well as the universal behavior near the critical point. Our results provide detailed description of critical thermodynamics near the BKT transition and offer new prospects to investigate other critical phenomena near classical or quantum phase transitions. In particular, we present experimental evidence of the growing correlations in the fluctuation region through the application of the fluctuation-dissipation theorem. Further investigations into the correlations will provide new insights into the rich critical phenomena near the transition point, for instance, critical opalescence and critical slowing.

\section*{Methods Summary}
Preparation and detection of cesium 2D Bose gases are similar to those described in Ref.~\cite{Hung10}
. We adjust the temperature of the sample by applying magnetic field pulses near a Feshbach resonance to excite the atoms. We then tune the scattering length to a designated value, followed by 800~ms wait time to ensure full thermalization of the sample.

Absorption imaging is performed \textit{in situ} using a strong resonant laser beam, saturating the sample to reduce the optical thickness. Atom-photon resonant cross-section and atomic density are independently calibrated. Averaged atom number $N_i$ and number fluctuation $ \delta N_i^2 $ at $i$-th CCD pixel are evaluated pixel-wise based on images taken under identical experiment conditions. The photon shot-noise, weakly depending on the sample's optical thickness, is calibrated and removed from the measured number variance. We correct the effect of finite imaging resolution on the remaining number variance using calibration from dilute thermal gas measurements. The density fluctuation $\delta n_i^2$ is obtained from the recovered atom number variance using $\delta n_i^2 \lambda_{dB}^2 = \delta N_i^2 /A$, which replaces the dependence on the CCD pixel area $A$ by a proper area scale $\lambda_{dB}^2$ (details see full Methods).


We thank Q. Zhou, B. Svistunov, T.-L. Ho, Y. Castin, C.-C. Chien, S. Tung, N. Prokof'ev, J. Freerick and D.-W. Wang for discussions. This work was supported by NSF (grant numbers PHY-0747907, NSF-MRSEC DMR-0213745), the Packard foundation, and a grant from the Army Research Office with funding from the DARPA OLE program. N.G. acknowledges support from the Grainger Foundation.


\section*{Methods}

\subsection{Calibration of the atomic surface density and the atom number fluctuation.}
Detection of caesium 2D Bose gases is detailed in ref. \cite{Hung10} and the atomic surface density $n$ of the 2D gas is evaluated with similar schemes discussed in Ref.~\cite{MethodReinaudi07}
, where the resonant cross-section $\sigma_0$ is independently calibrated using a thin 3D Bose condensate with similar optical thickness and the known atom number-to-Thomas-Fermi radius conversion. The calibrated value of $\sigma_0$ 
can be compared to that determined from the atom shot-noise amplitude in dilute 2D thermal gases, where the noise is evaluated using binned CCD pixels to remove finite resolution effects. For dilute thermal gases, we expect $\delta N^2  =  N $, where $N$ is the mean atom number; we compare the fluctuation amplitude to the mean and extract the value of $\sigma_0$. Two results agree to within 10$\%$ and the residual non-linearity in the density calibration is negligible.

We evaluate the atom number variance $ \delta N^2 $ pixel-wise based on images taken under identical experiment conditions. The photon shot-noise contribution $\delta N_p^2$, which weakly depends on the sample's optical thickness $n \sigma_0$, is calibrated and removed from the atom number fluctuation using $\delta N_p^2=  (\delta N_0^2/2)[1+\frac{(1+\gamma e^{-n \sigma_0})^2}{(1+\gamma)^2} e^{n\sigma_0}]$, where $\delta N_0^2$ is the photon shot-noise without atoms and $\gamma$ is the ratio of the imaging beam intensity to the saturation intensity. Both $\delta N_0^2$ and $\gamma$ are experimentally calibrated. We then correct for the effect of finite resolution on the number fluctuation\cite{Gemelke09} by comparing the atom number variance in a dilute thermal cloud to its mean atom number, using $\delta N^2= N$, 
and applying this calibration to all fluctuations measured at lower temperatures and higher densities.

\subsection{Density-density correlation in
the fluctuation measurement.} In the fluctuation measurement, we determine $\delta n^2$ from the pixel-wise atom number variance using the formula $\delta n^2 \lambda_{dB}^2 = \delta N^2 /A$, which replaces the dependence on the pixel area $A$ by a natural area scale $\lambda_{dB}^2$. This definition, however, does not fully eliminate the dependence on the imaging resolution spot size $v \sim (2~\mu$m$)^2$. In particular, when the density-density correlation length $\xi$ approaches or exceeds the resolution, the measured fluctuation can depend on the fixed length scale $\sqrt{v}$, which can complicate the scaling behavior. However, we do not see clear deviation of scale invariance and universality within our measurement uncertainties (Fig.~2b and 3b). We attribute this to the small variation of the non-scale invariant contribution within our limited range of sample temperature. Further analysis on the correlations and fluctuations is in progress and the result will be published elsewhere. 

\subsection{Determination of the BKT critical values from the fluctuation data.} We use a hyperbolic function $y(\tilde{\mu})=s (\tilde{\mu}-\tilde{\mu}_c)-\sqrt{s^2(\tilde{\mu}-\tilde{\mu}_c)^2+w^2}$ to empirically fit the crossover feature of the density fluctuation near the transition region, assuming $ \delta \tilde{n}^2(\tilde{\mu}) =D e^{y(\tilde{\mu})}$, where the critical chemical potential $\tilde{\mu}_c$, the fluctuation in the superfluid regime $D$, the slope of the exponential rise $s$, and the width of the transition region $w$ are fitting parameters. The critical phase space density is then determined from the density profile as $\tilde{n}_c=\tilde{n}(\tilde{\mu}_c)$. Other choices of fit functions give similar results, contributing only small systematics from the choice of different models.

\subsection{Obtaining the universal function $H(x)$.} We use the density profiles in the inset of Fig.~3a to look for critical values $\tilde{n}_c$ and $\tilde{\mu}_c$ such that the equations of state at all values of $g$ collapse to a single universal curve $H(x)=\tilde{n}(\tilde{\mu})-\tilde{n}_c$, where $x=(\tilde{\mu}-\tilde{\mu}_c)/g$ is the rescaled chemical potential. To do this, we take the profile measured at $g=0.05 \equiv g_r$ as the reference, evaluate $H_r(x)=\tilde{n}(g_r x+\tilde{\mu}_{c,r})-\tilde{n}_{c,r}$ using the critical values $\tilde{n}_{c,r}$ and $\tilde{\mu}_{c,r}$ determined from the fluctuation crossover feature, and smoothly interpolate the data to make a continuous reference curve $H_r(x)$ in the range of $|x|\leq 1$. Using this model, we perform minimum chi-square fits to the profiles measured at all other values of $g$ according to $\tilde{n}(\tilde{\mu})=\tilde{n}_c+H_r(\frac{\tilde{\mu}-\tilde{\mu}_c}{g})$, with only $\tilde{n}_c$ and $\tilde{\mu}_c$ as free parameters. This procedure successfully collapses all density profiles (see Fig.~3a), and is independent of any theoretical model. The resulting critical values $\tilde{n}_c$ and $\tilde{\mu}_c$ are plotted in Fig.~3c-d. 


\end{document}